\documentclass{aa}  
\usepackage{graphicx}
\usepackage{txfonts}
\usepackage{natbib}
\usepackage{subeqn}
\newcommand{\msun}{\ensuremath{\mathit{M}_{\odot}}}                  
\newcommand{\lsun}{\ensuremath{\mathit{L}_{\odot}}}
\newcommand{\rsun}{\ensuremath{\mathit{R}_{\odot}}}
\newcommand{\msunyr}{\ensuremath{\mathit{M}_{\odot}{\rm yr}^{-1}}}   
\newcommand{\meanP} {$\langle P \rangle $}

\newcommand{\kms} {~km~s$^{-1}$}

\begin{document}
   \title{Modelling the clumping-induced polarimetric variability \\ of hot star winds}

   \author{Ben Davies\inst{1,2}
          \and
          Jorick S. Vink\inst{3,4,5}
	  \and
	  Ren\'{e} D. Oudmaijer\inst{2}
          }

   \offprints{B. Davies, {\tt davies@cis.rit.edu}}

   \institute{Center for Imaging Science, Rochester Institute of Technology, 84
     Lomb Memorial Drive, Rochester, NY 14623, USA
         \and
	 School of Physics \& Astronomy, University of Leeds, 
	 Woodhouse Lane, Leeds LS2 9JT, UK
	 \and
	 Imperial College, Blackett Laboratories, Prince Consort Road,
	 London SW7 2BZ, UK
	 \and
         Lennard-Jones Laboratories, Keele University, Staffordshire
	 ST5 5BG, UK
	 \and
	 Armagh Observatory, College Hill, Armagh BT61 9DG, NI, UK
   }

   \date{Received ; accepted }

 
  \abstract {Clumping in the winds of massive stars may significantly
reduce empirical mass-loss rates, and which in turn may have a large
impact on our understanding of massive star evolution.}  {Here, we
investigate wind-clumping through the linear polarization induced by
light scattering off the clumps.} {Through the use of an analytic wind
clumping model, we predict the time evolution of the linear
polarimetry over a large parameter space. We concentrate on the
Luminous Blue Variables, which display the greatest amount of
polarimetric variability and for which we recently conducted a
spectropolarimetric survey.} {Our model results indicate that the
observed level of polarimetric variability can be reproduced for two
regimes of parameter space: one of a small number of massive,
optically-thick clumps; and one of a very large number of low-mass
clumps.}  {Although a systematic time-resolved monitoring campaign is
required to distinguish between the two scenarios, we currently favour
the latter, given the short timescale of the observed polarization
variability.  As the polarization is predicted to scale linearly with
mass-loss rate, we anticipate that all hot stars with very large
mass-loss rates should display polarimetric variability. This is
consistent with recent findings that intrinsic polarization is more
common in stars with strong H$\alpha$ emission.}

   \keywords{Stars: mass-loss -- Stars: winds, outflows -- Polarization}

   \maketitle
%

\section{Introduction}
Many different species of massive star display intrinsic linear
polarization -- from main-sequence O stars \citep[][ LN87
  hereafter]{L-N87}, to post main-sequence transitional objects such
as OB supergiants \citep[][ LN87]{Hayes84}, Red Supergiants
\citep{Hayes84rsg}, Luminous Blue Variables \citep[LBVs, ][]{Davies05}, B[e]
supergiants \citep[]{Melgarejo01} and Wolf-Rayet stars \citep[WRs,
][]{StL87}. This intrinsic polarization results from the scattering of
starlight by aspherical structure in the stellar wind. In the case of
the B[e]SGs, this inhomogeneity is thought to be in the form of a
circumstellar disk, although the data for this is limited
\citep[e.g.][]{Z-SL89,S-L93}. The intrinsic polarization of
the LBV \object{AG Carina} was also thought to result from a global
axi-symmetric wind structure \citep{S-L94}. However, multi-epoch
observations have shown that the polarization {\it angle} of \object{AG
Car} is stochastically variable \citep{Davies05}. The polarization of
\object{P Cygni} was found to be variable on timescales of $\la$ weeks
\citep{Nordsieck01}, which is similar to the short timescale
variability observed in WRs \citep{Robert89}. This behaviour is
commonly explained as being due to wind-clumping (e.g. LN87).

Not just a quirk of massive stars, wind-clumping could have serious
implications for our understanding of stellar evolution. Mass-loss
rates have traditionally been determined by fitting recombination
lines with models that assume a smooth outflow \citep[for LBVs see
e.g.][]{Stahl01,Machado02}. As recombination is a two-body interaction
process it depends on $\rho^2$, so inhomogeneity will result in an
over-estimation of the density {\bf (as $\langle \rho^{2} \rangle / \langle
\rho \rangle ^{2}$)}, and hence an over-estimation of the mass-loss
rate.

Recently, there has been much discussion about clumping in the winds
of O stars.  The value of the clumping factor $f_{\rm cl}~=~ \langle
\rho^{2} \rangle / \langle \rho \rangle ^{2}$ is as yet an open
issue. On the one hand, H$\alpha$ analyses by \citet{repolust04} and
\citet{mokiem07} find that a modest amount of wind clumping, with a
clumping factor up to $f_{\rm cl} \simeq 5$ is required to match the
smooth line-driven wind models of \citet{Vink00}. But on the other
hand, X-ray \citep[e.g.][]{oskinova06} and quantitative UV studies
\citep[e.g.][]{Bouret05,Fullerton06} call for $f_{\rm cl}$ values much
larger than 10.  If confirmed, this could significantly affect our
understanding of massive star evolution.

In addition to spectral diagnostic techniques, linear polarimetry
studies have proven to be a powerful tool to probe the geometry of the
wind structure. For instance, the results of a spectropolarimetric
survey of LBVs by \citet{Davies05} call for an onset of wind clumping
in close proximity to the stellar photosphere, as to be able to
reproduce the observed levels of polarization.  LBVs may be an ideal
testbed to study hot star wind clumping as the strength of the
intrinsic polarization is much larger than that observed in O and WR
stars. Systematic monitoring campaigns have shown that WRs display
polarization around the 0.1\% level, with variability on very short
timescales of $<$ 1 day. Additionally, the variability was found to
have a greater amplitude for those stars with slower terminal wind
speeds \citep[ $v_{\infty}=1200 \rightarrow 3600 {\rm km s}^{-1}$,
$\Delta P \sim 0.1 \rightarrow 0.02\%$; ][]{Robert89}. LBVs have wind
speeds around a factor of 10 slower and show intrinsic polarization of
$\ga$0.5\%. Although there have yet to be any systematic
polarimetric monitoring campaigns of LBVs, the data of \citet{Hayes85}
appeared to indicate variability on timescales comparable to the wind
flow-time ($t_{\rm fl} \equiv R_{\star}/v_{\infty}$), which for LBVs
is on the order of $\sim$days. This is consistent with the
polarization arising within a few stellar radii of the central star
\citep[c.f.\ ][]{Cassinelli87}.


Theoretical studies of clumping-induced polarization are few, and
tend to be `first-step' investigations rather than comprehensive
studies. \citet{R-M00} and \citet{C-W95} presented Monte-Carlo studies
of the polarization produced by a single clump. \citet{Harries00}
studied the time-variability of a clumpy wind, again with a
Monte-Carlo code, but did not investigate the effect of changing
stellar/wind properties, nor the polarimetric variability on longer
timescales. Both of these studies concluded that the clumps must be
optically thin, as increasing the clump optical depth begins to have a
conspicuous effect on time-resolved observations of spectral
line-profiles. \citet{Li00} used an analytical model to investigate
the polarimetric variability of WRs, but again did not study the
effect of different stellar parameters. \citet{Richardson96} studied
specifically the effect of a clumpy wind on the ratio of photometric
to polarimetric variability in the case of WRs. \citet{Brown00}
investigated the `redistribution' of a smooth outflow into clumps,
simulating the pile-up of material caused by e.g.\ shocks in the
wind. They concluded that small-scale redistribution close to the star
{\it could not} produce significant levels of polarization. This
implies that clumps producing short-timescale polarimetric
variability, which are necessarily close to the star, must be
localized over-densities of ejecta, as opposed to the result of a
smooth outflow being `snow-ploughed' by a shock-front.

In this paper we use an analytic model to build on these previous
studies and fully investigate the polarimetric variability produced by
a clumpy wind across a range of stellar parameters typical of massive
evolved stars. We place particular emphasis on LBVs -- the group of
stars which display the strongest polarimetric variability. A full
description of the model is presented in Sect.\,\ref{sec:modeldesc},
along with a few basic applications to geometries explored in other
work to serve as validation. The relevant input parameters are
discussed in Sect.~\ref{sec:parameters}, leading to the model results
and discussion in Sects.~\ref{sec:modelresults}
and~\ref{sec:modeldisc}.



\section{Description of the model} \label{sec:modeldesc}
In this analytic study we follow the approach of, amongst
others, \citet{B-M77,Brown00,Li00}. The strategy for the construction
of the model is as follows: the polarization due to a single clump is
determined as a function of distance from the illuminating
star. Through the velocity-law of the wind, an expression for the
polarization as a function of time is found. Multiple clumps are
added, and the total polarization found from the vector sum of the
polarization of each individual clump. The clumps are then given a
typical ejection timescale, and the polarization as a function of time
is calculated as new clumps are ejected and move through the
wind. After the simulation reaches a steady-state, the
polarization is monitored for a further period of time to determine
the typical polarization level and its variability.


%

\subsection{Model geometry}

\begin{figure}[h]
  \centering
  \includegraphics[width=8cm,bb=0 0 380 280,clip]{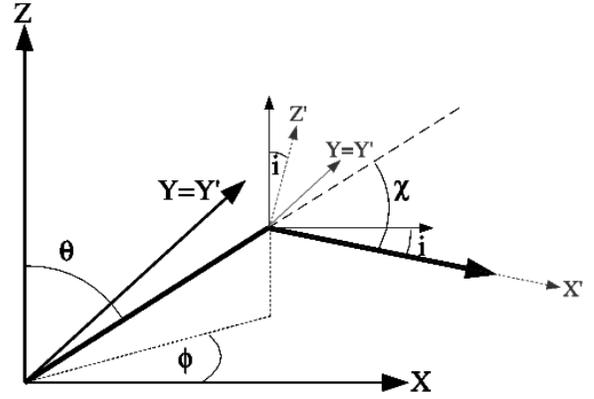}
  \caption{The coordinate system used in this model, as defined in
  \citet{B-M77}. A beam with direction ($\theta,\phi$) scatters
  through an angle $\chi$ towards the observer. The observer's
  reference frame is rotated through an angle $i$ with respect to the
  star's reference frame. \label{fig:BMgeom}}
\end{figure}

The coordinate system used is that defined in \citet{B-M77}, and is
illustrated in Fig.\,\ref{fig:BMgeom}. The angle of the plane of
scattering to the observer $\chi$ can be described in terms of the
polar ($\theta$) and azimuthal ($\phi$) angles of the star's reference
frame, and the inclination of this frame to the plane of the sky $i$:

\begin{equation}
\cos\chi = -\cos \theta \cos i + \sin \theta \sin i \cos \phi
\label{equ:scatter}
\end{equation}

\noindent  The position angle (PA) of the polarization, $\Theta$ , is
described by,

\begin{equation}
\sin {\Theta} = \frac{ \cos\theta\sin i + \sin\theta\cos i\cos\phi}
{\{(\cos\theta\sin i+\sin\theta\cos
i\cos\phi)^{2}+\sin^{2}\theta\sin^{2}\phi\}^{0.5}}
\label{equ:pa}
\end{equation}

\subsection{Polarization of axi-symmetric material}
\begin{figure}[h]
  \centering
  \includegraphics[width=8cm,clip]{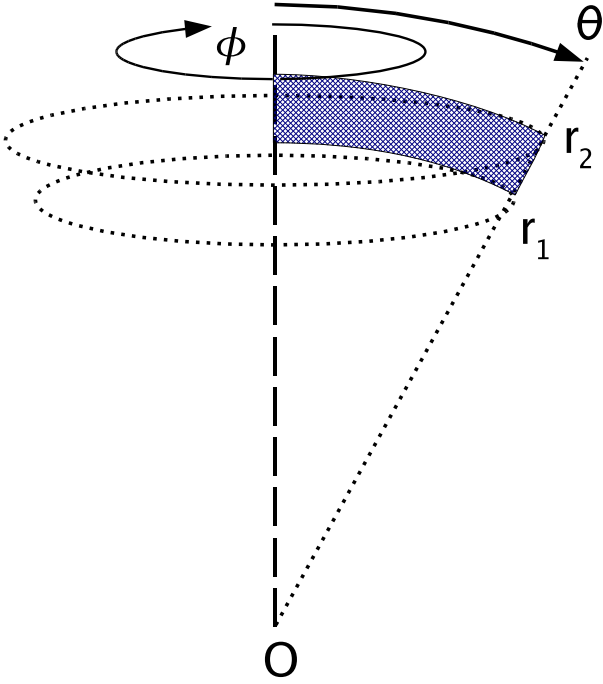}
  \caption{The geometry assumed in
    Eq.~(\ref{equ:pol}). The electron density is assumed to be symmetric
    about the central axis (i.e. over all $\phi$), such that by
    specifying the limits $r_{\rm 1}$ to $r_{\rm 2}$ and $\theta_{\rm
    1}$ to $\theta_{\rm 2}$ we are integrating over the volume swept
    out by the shaded area, in this case a `shell'.
    \label{fig:coord_geom}}
\end{figure}

From \citet{Brown00}, which uses the expressions of \citet{B-M77} and
\citet{Cassinelli87}, the polarization produced by electron scattering
material in a stellar wind as a function of distance to the star $r$,
polar angle $\theta$, and azimuthal angle $\phi$, is described by,

\begin{eqnarray*}
P(\mu,r)=\frac{3}{16} \sigma_{\rm T} \sin^{2} \chi
  \displaystyle\int^{\mu_{2}}_{\mu_{1}} (1-3\mu^{2}) d\mu \\ \times
  \int^{r_{2}}_{r_{1}} n_{\rm e} (r)
  \sqrt{1-\left(\frac{R_{\star}}{r}\right)^2}dr
\label{equ:pol}
\end{eqnarray*}

\noindent where $\sigma_{\rm T}$ is the Thompson scattering
cross-section, $\chi$ is the angle between the plane of scattering and
the observer, $\mu = \cos \theta$, $R_{\star}$ is the radius of the
star, and $n_{\rm e}$ is the electron number density which in this
case is a function of $r$ only. The square-root factor within the $r$
integral is known as the {\em finite-star correction-factor}
\citep{Cassinelli87}, and accounts for the finite stellar `disk' seen
by the scattering material.  This formalism can be used to determine
the polarization produced by different scattering geometries, and an
illustration of this can be seen in Fig. \ref{fig:coord_geom}.

\subsection{Clumps} \label{sec:clumps}
The amount of polarization a clump produces is determined by how much
light is intercepted, as well as the clump's size and
morphology. Arbitrary changes in clump morphology can produce minor
changes in $P$ {\it if} the clump is close to the star, due to the
effect of the finite size of the illuminating star. This effect is
small, however, and it is a reasonable approximation to say that, for
a constant column-density, halving a clump's size (and hence doubling
its density) produces no difference in polarization \citep[see
also][]{Brown00}. For this reason we concentrate on the simplest
morphology for the clumps, that of constant angular size and
thickness. In a radiatively driven wind this scenario is plausible, as
each particle in the wind is driven radially outwards, and any
inhomogeneity will grow in size with the square of the distance.


\subsubsection{Polarization of a single `shell' clump}
The geometry of Fig. \ref{fig:coord_geom} and Eq. (\ref{equ:pol}) are
used as a starting point. The angular size remains
constant with distance from the star, so $\mu$ can be integrated from
$\mu_{1} = 0$ to $\mu_{2} = 1-\Delta\Omega/2\pi$, where $\Delta\Omega$
is the angular size of the clump. 
The electron number density is parameterized in order to conserve mass
for a clump with constant angular size, thickness and number of
electrons, i.e.

\begin{equation}
n(r) = n_{0}\displaystyle\left(\frac{R_{\star}}{r}\right)^2
\label{equ:dens}
\end{equation}
  
\noindent where $n_{0}$ in the initial number density of the clump at
$r = R_{\star}$. 

Substituting Eq. (\ref{equ:dens}) into Eq. (\ref{equ:pol}) and 
setting $x = r/R_{\star}$, the polarization of a single clump at a
distance $x'$ from a star is,

\begin{eqnarray*}
P & = & \frac{3}{16}\sigma_{T}n_{0}R_{\star} ~
\sin^2\chi ~ (\mu_{2}-\mu_{2}^{3})\\ && \times
\int^{x'+\Delta r}_{x'} \frac{1}{x^{2}} \sqrt{1-\frac{1}{x^2}}~~dx
\label{equ:pshell}
\end{eqnarray*}

\noindent where $\Delta r$ is the dimensionless thickness of the
shell.

As this is an analytical study, multiple scattering is not taken into
account in this equation. 
When multiple scattering effects start to become relevant,
i.e. when $\tau_{e} > 1$, the results from Eq.~(\ref{equ:pshell}) would become 
unphysical. To assess the importance of multiple scattering, we need 
to monitor the optical depth, and we re-write Eq. (\ref{equ:pshell}) 
in terms of the initial optical depth per clump 
$\tau_{0}$, i.e. when $r = R_{\star}$:

\begin{eqnarray*}
\tau_{0} & = & \frac{3}{16}\sigma_{T} \int_{r_{1}}^{r_{2}}n(r)~dr \\
& = & \frac{3}{16}\sigma_{T}n_{0}R_{\star}K
\end{eqnarray*}

\noindent where,

\begin{eqnarray*}
K & = & \int^{1+\Delta r}_{1}\frac{1}{x^{2}}~dx \\ 
& = & 1 - \frac{1}{1+\Delta r} \\ \\
& \simeq & \Delta r ~~ ({\rm if}~\Delta r \ll 1)
\label{equ:K}
\end{eqnarray*}

In order to relate the initial optical depth to other observables, the
initial optical depth per clump can be expressed in terms of the
stellar parameters. The mass-loss rate and clump ejection rate, $\dot{N}$, 
are related via,

\begin{equation}
 \dot{M} = \dot{N} N_{e} \mu_{e} m_{H}
\end{equation}

\noindent where $N_{e}$ is the number of electrons in each clump,
$\mu_{e}$ is the mean mass-per-particle and $m_{H}$ is the mass of a
proton. To parameterize the ejection rate in the general case, the
formalism of \citet{Li00} is adopted by defining the ejection rate per
{\it wind flow-time} $\mathcal{N}$, where $t_{\rm fl} \equiv
R_{\star}/v_{\infty}$:

\begin{equation}
{\mathcal N} = \dot{N} t_{\rm fl} = \dot{N} R_{\star} / v_{\infty}
\end{equation}

\noindent Therefore, the number of electrons in each clump:
\begin{equation}
N_{e} = \frac{\dot{M} R_{\star}} {\mu_{e} m_{H} ~ {\mathcal N} v_{\infty} }
\end{equation}

\noindent The initial volume per clump is,

\begin{equation}
V_{\rm cl} = \Delta\Omega R_{\star}^{3} \Delta r
\end{equation}

\noindent where $\Delta\Omega$ is the solid angle $2\pi(1-\mu_2)$, and
$\Delta r$ is the clump thickness in units of $R_{\star}$ and is
assumed to be $\ll$1. The electron
density per clump is then,
 
\begin{equation}
n_{0} = \frac{N_{e}}{V_{\rm cl}} =
\frac{\dot{M}}{\mu_{e} m_{H} ~ {\mathcal N} v_{\infty} ~ \Delta\Omega
  R_{\star}^{2} \Delta r}
\label{equ:n0}
\end{equation}

\noindent Equation (\ref{equ:pshell}) can now be re-written as,

\begin{eqnarray*}
P & = & \frac{\tau_{0}}{\Delta r} ~ \sin^2\chi ~ (\mu_{2}-\mu_{2}^{3})\\ && \times
\int^{x'+\Delta r}_{x'} \frac{1}{x^{2}} \sqrt{1-\frac{1}{x^2}}~~dx
\label{equ:pshell2}
\end{eqnarray*}

\noindent where,

\begin{equation}
\frac{\tau_{0}}{\Delta r} = \frac{3}{16}\sigma_{T} ~
\frac{\dot{M}}{\mu_{e} m_{H} ~ {\mathcal N} v_{\infty} ~ \Delta\Omega R_{\star} \Delta r}
\label{equ:tau0}
\end{equation}


\begin{figure}[h]
  \includegraphics[width=8cm]{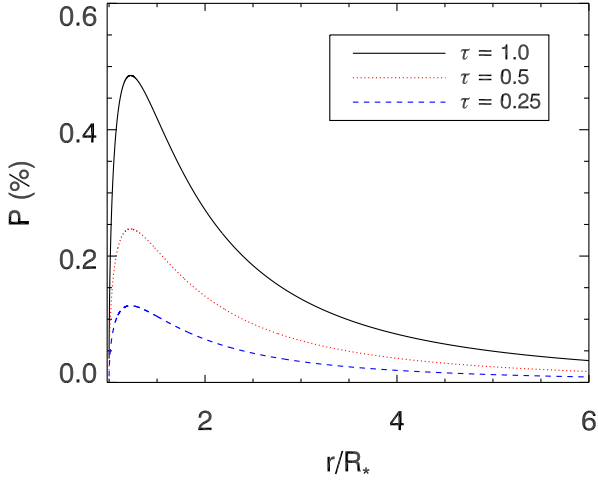}
  \caption{The polarization as a function of distance for a single
  clump for three values of $\tau$. The clump size is
  $\Delta\Omega=0.04, ~\Delta r = 0.01$; and $\sin^{2} \chi = 1$. The
  polarization falls-off as roughly 1/$r$, with the attenuation at
  small $r$ due to the finite size of the illuminating star. It can be
  seen that at distances of $r/R_{\star} \sim 5$ the polarization has
  already decreased from the maximum value by a factor of $\sim$10.}
  \label{fig:oneclumpx}
\end{figure}

\noindent The results of the polarization of a single clump as a function of 
distance are shown for 
different optical depth values $\tau$ in Fig.~\ref{fig:oneclumpx}.

\subsubsection{Polarization of a single clump as a function of time}
Having determined the polarization as a function of distance from the
central star, we calculate the time behaviour of the polarization. 
The velocity of the wind as a function of distance from the star is
described by the so-called `beta-law',

\begin{equation}
v(r) = v_{\infty}\displaystyle\left(1-\frac{bR_{\star}}{r}\right)^{\beta}
\label{equ:betalaw}
\end{equation}

\noindent where $v_{\infty}$ is the terminal velocity of the wind, $b$
is a constant which dictates the initial velocity of the wind $v_{0}$
(i.e. at $r=R_{\star}$), and $\beta$ is a constant which determines
the wind's acceleration. Hot supergiants are typically thought to have
$1 < \beta < 2$ \citep[see e.g.][]{Herrero02}, so for ease of
calculation a value of $\beta = 1$ is used throughout most of this
study. By rearranging Eq.\,(\ref{equ:betalaw}), setting $\beta=1$, and
substituting $x = r/R_{\star}$, we find an expression for the time
since ejection $t$ as a function of distance $x$ and the velocity-law
parameters $v_{\infty}$, $\beta$, and $b$,
 
\begin{equation}
t = \int \frac{{\rm d}r}{v(r)} = \frac{R_{\star}}{v_{\infty}}[x+b\ln(x-b)+c]
\label{equ:tpol}
\end{equation}

\noindent where $c$ is a constant of integration, which is determined
by forcing $x=1$ at $t=0$. The polarization as a function of time can
now be calculated by solving Eqs. (\ref{equ:pshell}) and
(\ref{equ:tpol}) parametrically.  The initial velocity $v_{0}$ is set
to the isothermal sound speed at the temperatures considered. Altering
$v_{0}$ by a factor of two was found to have little difference on the
results of a given simulation, so for simplicity the value $v_{0} =
10$\kms\ is used throughout this study.

\begin{figure}[t]
  \centering
  \includegraphics[width=8cm]{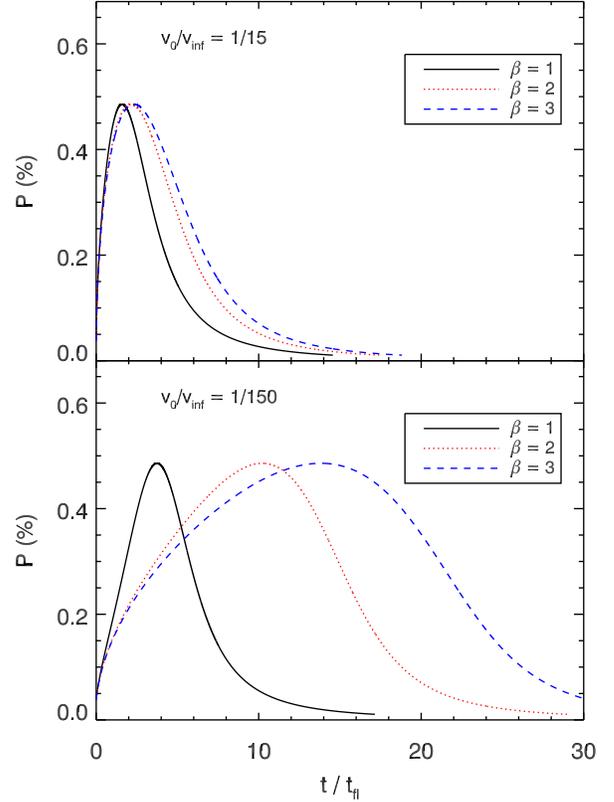}
  \caption{The polarization as a function of time (in units of wind
  flow-time) for three different values of $\beta$. {\it Top}: for
  slow terminal wind speed; {\it Bottom}: fast terminal wind
  speed. Same clump parameters are used as for
  Fig.\,\ref{fig:oneclumpx}. }
\label{fig:oneclumpt}
\end{figure}

The effect of increasing $\beta$ is shown in
Fig.\,\ref{fig:oneclumpt}. The plot shows the polarization as a
function of time for three values of $\beta$, and two separate ratios
of $v_{0}/v_{\infty}$. As increasing $\beta$ increases the time the
clump spends in the inner regions of the wind (the `dwell time'), the
polarization remains high for longer. The effect is more pronounced
when $v_{0}/v_{\infty}$ is higher.



\subsubsection{Polarization of multiple clumps}
Clumps are ejected radially from a random location on the stellar
surface, with $0 < \phi_{j} < 2\pi$ and $-1 < \cos\theta_{j} < 1$ for
the $j$th clump. For a given inclination angle, the polarization and
its position angle due to a single clump as a function of time are
calculated through Eqs. (\ref{equ:scatter}), (\ref{equ:pa}),
(\ref{equ:pshell2}), and (\ref{equ:tpol}). Clumps are ejected over a
typical timescale $T_{\rm ej}$, such that $T_{\rm ej} =
R_{\star}/\mathcal{N} v_{\infty}$. The effect of randomizing $T_{\rm
  ej}$ was experimented with, such that $T_{\rm j} = T_{\rm ej} \pm
\sigma_{\rm ej}$. This was found to have no effect on the results,
and was computationally more expensive as it required finer grids to
be set up. Consequently, a uniform   $T_{\rm ej}$ was used throughout
each simulation.

To calculate the total polarization due to all clumps at any one time,
the polarization due to each clump must be converted into the linear
Stokes parameters,

\begin{subequations}
\begin{equation}
Q = P \cos 2\Theta
\end{equation}
\begin{equation}
U = P \sin 2\Theta
\end{equation}
\end{subequations}

\noindent where $\Theta$ is perpendicular to the plane of
  scattering. The total polarization at any one time is then
  determined from the sum of the Stokes vectors of each of the $N$
  clumps in the wind,

\begin{subequations}
\begin{equation}
P_{\rm tot}^2 = \displaystyle \left( \sum_{j=1}^{N} Q_{j} \right)^2
+ \left( \sum_{j=1}^{N} U_{j} \right)^2
\end{equation}
\begin{equation}
\Theta_{\rm tot} = \frac{1}{2} \tan^{-1} \displaystyle \left(
\sum_{j=1}^{N} U_{j} / \sum_{j=1}^{N} Q_{j} \right)
\end{equation}
\end{subequations}

In order to determine the typical polarization of a model run, the
simulation is run until it reaches an equilibrium, and the first
clumps to be ejected no longer contribute significantly to the overall
polarization. This occurs when the first clumps reach $r/R_{\star} \ga
6$ (see Fig. \ref{fig:oneclumpx}).  The polarization is then still
monitored for a further 30 flow-times, and the time-averaged
polarization \meanP\ over this period, together with the
r.m.s. variability $\Delta P$, are recorded. As the simulation is
stochastic by design, it can sometimes throw up atypical results. To
compensate for this, the process is repeated several times for each
set of parameters, and the mean is taken. The experimental
uncertainties in \meanP\ and $\Delta P$ were taken to be $\sigma /
\sqrt{n}$, where $\sigma$ is the r.m.s. scatter of the results of $n$
simulations.

\subsubsection{High ejection-rates}
The goal of this work is to reproduce the polarimetric variability 
of hot stars -- both on long timescales, as found by the repeated
measurements in the literature; and short-timescales, as found by
e.g. \citet{Hayes84} and \citet{StL87}. At low ejection rates,
individual ejection events may be resolvable by nightly monitoring,
and the model would have to be {\it dynamic} to simulate this,
i.e. follow the polarimetric behaviour in real-time as clumps move
through the wind, whilst new clumps are ejected.

However, at high ejection rates the distribution of clumps close to 
the photosphere will look completely different from night-to-night, and 
so will the polarization. In this situation, there is nothing to be
gained from computationally-expensive dynamical modelling as ejection
events are unresolvable. In the high-$\mathcal{N}$ regime,
calculations are therefore performed using a {\it static} model. 
Here, as the polarization at each observation is unrelated to the last,
a series of wind `snap-shots' are generated, which represent a sample
of independent observations. Clump distances are randomly-generated
according to the ejection rate and velocity law, and the total
polarization is calculated. This is repeated $10^{3}$ times, and
$\bar{P}$ and $\Delta P$ calculated.

Whilst being unable to predict the time-evolution of individual
clumps, this static method has the advantage of being able to 
simulate a much broader range of $\mathcal{N}$ values -- at drastically 
reduced computation times.

\subsection{Optically-thick clumps}
Equation (\ref{equ:pol}) shows that, in this study, the amount of
polarization produced by a clump is proportional to the clump's
optical depth, as denser clumps scatter more light.
 
However, this will only hold as long as the clump is optically thin.
When the optical depth becomes larger than $\sim 1$, multiple
scattering becomes important. In this regime, the relation in
Eq.\,(\ref{equ:pol}) breaks down as a photon which is scattered twice
cannot produce twice as much polarization.

To investigate the effect of optically-thick clumps, a first-order
approximation is made from the Monte-Carlo work of \citet{R-M00}. They
studied the polarization produced by starlight scattering off clumps
with optical depths as high as $\tau_{\rm e} = 10$. Their results show
that, in the low-$\tau_{\rm e}$ regime, the polarization rises almost
linearly with optical depth, until $\tau$ reaches $\sim 1$. At
high-$\tau_{\rm e}$ the polarization flattens off. This can be
understood in terms of multiply-scattered photons `forgetting' where
they came from, with only the singly-scattered photons carrying the
scattering geometry information. This explains the slight fall-off of
$P$ with high $\tau$, as less and less of the observed photons are
singly-scattered.

We approximate the results of \citet{R-M00} such that
polarization is linear with optical depth up to $\tau = 1$, and then
constant for higher $\tau$. In this situation, as an optically-thick
clump moves away from the star its density decreases as $1/r^{2}$, but
the polarization remains constant while the optical depth is
larger than 1. When the clump reaches a distance such that $\tau =
1$, its polarization decreases rapidly in the same manner as an
optically-thin clump. 

In making this approximation, we note that both \citet{C-W95} and
\citet{R-M00} considered the case of {\it spherical} clumps, while in
this work we consider only flattened, `shell' clumps. Although we
argue in Sect.\ \ref{sec:clumps} that clump shape is a redundant
parameter in the optically-thin limit, the relative invariance of
clump morphology to polarization may not hold for optically-thick
clumps. For example, in the case of a spherical clump, the limb will
always be optically-thin to some extent, whereas the optical depth of
a `shell' clump will always be the same along all radial paths from
the star. This problem bears some similarity to the challenges
faced when modelling clouds in planetary atmospheres
\citep{vdHulst80}. The impact and limitations of our approximation in
the context of the results will be discussed further in
Sect.~\ref{sec:resthick}.


\section{Model parameters}
\label{sec:parameters}

For a given mass-loss rate and ejection timescale, the key input
parameters are: the terminal wind velocity, the clump size and optical
depth, the stellar radius, and the ejection timescale.  Each of these
parameters are discussed below.

\subsection{Clump size}
The size of each clump is determined by its angular size
$\Delta\Omega$ and its thickness $\Delta r$.  Provided $\Delta r$ is
small, it has no effect on the polarization per clump as the optical
depth remains the same.
Varying $\Delta\Omega$ has a small effect, as for large $\Delta\Omega$
the polarization at the edge of a shell slightly cancels that of the
opposite side. This can be seen in the simulations
(Sect. \ref{sec:results_size}).

\subsection{Terminal wind velocity $v_{\infty}$, and stellar radius
$R_{\star}$} The polarization produced by a single clump falls off
quickly with distance to the star (see Fig. \ref{fig:oneclumpx}). The
amount of time that each clump will spend close to the star (the
`dwell time') is therefore a major factor in the overall level of
polarization. The dwell time is affected not only by speed of the
outflow, but also the size of the star. This is therefore reflected in
the star's wind flow-time $t_{\rm fl} \equiv R_{\star}/v_{\infty}$.

The terminal velocity is typically a few times the stellar
escape velocity, which in turn depends on the effective gravity of the
star:

\begin{equation}
g_{\rm eff} = g_{N} - g_{\rm rad} 
\end{equation}

\noindent where $g_{\rm N}$ is the Newtonian gravity $GM/R_{\star}^2$
and $g_{\rm rad}$ is the outward radiative acceleration due to the
continuum opacity of free electrons in the ionised stellar atmosphere,

\begin{equation}
g_{\rm rad} = \frac{\kappa_{e} L_{\star} }{4 \pi R_{\star}^2 c}
\end{equation}

\noindent The effective mass is then defined as,

\begin{equation}
M_{\rm eff} = (1 - \Gamma_{\rm e}) M_{\star}
\end{equation}

\noindent where $\Gamma_{\rm e} = g_{\rm rad} / g_{\rm N}$. The
terminal wind velocity is now,

\begin{eqnarray}
v_{\infty} &=& Q_{\rm BSJ} v_{\rm esc} \nonumber \\
&=&  Q_{\rm BSJ} \sqrt{ \frac{2GM_{\rm eff}}{R_{\star}} }
\label{equ:vinf_bsj}
\end{eqnarray}
\label{sec:bsj}

\noindent where $Q_{\rm BSJ}$ is the ``bi-stability jump'' ratio which is 
predicted to depend on the star's effective temperature
\citep{Vink99}. According to empirical studies \citep{Lamers95,K-P00}, 
$Q_{\rm BSJ}$ takes two distinct values,

\begin{eqnarray}
Q_{\rm BSJ} \simeq 1.3 & (T_{\rm eff} \la 21000{\rm K}) \nonumber \\
Q_{\rm BSJ} \simeq 2.5 & (T_{\rm eff} \ga 21000{\rm K}) \nonumber
\end{eqnarray}

\noindent As the stellar radius and effective temperature are related
via,

\begin{equation}
L_{\star} = 4 \pi R_{\star}^2 \sigma_{\rm B} T_{\rm eff}^{4}
\end{equation}

\noindent the terminal wind velocity can be found from the star's
mass, luminosity and temperature.  In the case of LBVs, this allows us
to predict how the terminal wind-speed varies over the S~Doradus
phases which are thought to occur at constant bolometric luminosity.


We note that a recent study of blue supergiants over the bi-stability
regime indicates that $Q_{\rm BSJ}$ may not be a step-function, but
instead it may decrease smoothly with decreasing temperature
\citep{Crowther06}. The binning of the data by effective temperature
by Lamers et al.\,may give the impression that $Q_{\rm BSJ}$ behaves
like a step function.
One of the main reasons $Q_{\rm BSJ}$ may not be a step function is
that one is comparing objects with very different stellar
parameters. For {\it individual} objects, as is relevant for our
study, the wind velocity may nonetheless be expected to 'flip-flop',
especially in the domain of LBVs \citep[e.g.][]{P-P90,V-dK02}.

\section{Results} \label{sec:modelresults}
In the first part of this section, we compare our model to the results
of a similar study by \citet{Li00}. Next, the effects of the various
input parameters are discussed one-by-one. By constraining various
stellar parameters to those of the well-studied,
polarimetrically-variable P Cyg, the more detailed parameter-space is
explored.

\subsection{Comparison with previous work} \label{sec:comparison}
\begin{figure}[t]
\centering
\includegraphics[width=8.5cm,bb=20 20 470 470,clip]{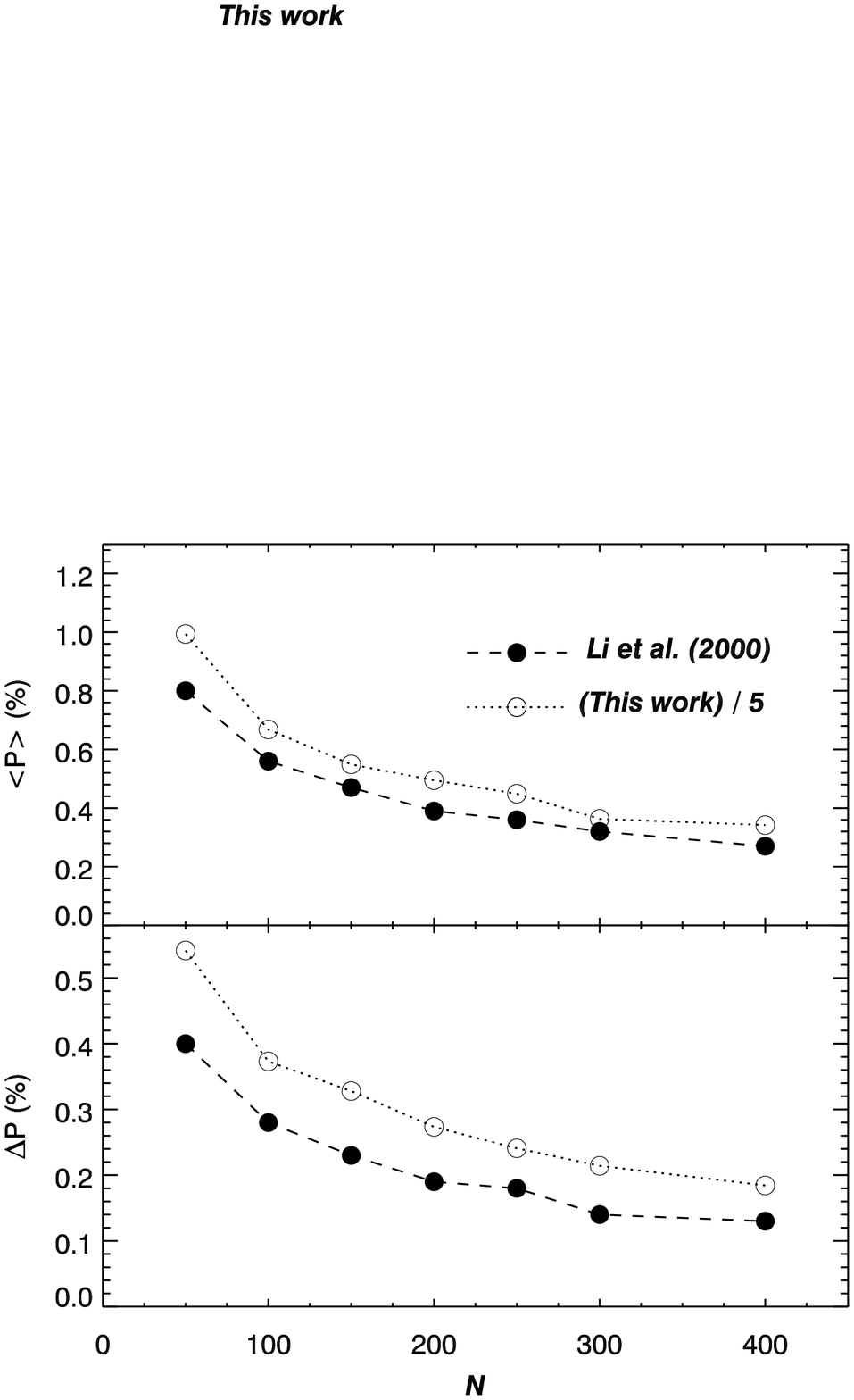}
  \caption[Comparison between results of Li et al.~2000 and this
  work.]{Comparison between the results of \citet{Li00} and the
  results of this work {\it when divided by a scaling factor of
  5}. The top panel shows the time-averaged polarization as a
  function of clump ejections per wind flow-time, whilst the bottom
  panel shows the variability in $P$. The results of the present model
  have been scaled to highlight the fact that while
  the two models agree qualitatively, quantitatively they differ. 
  Parameters used are $v_{\infty} = 1800$km s$^{-1}$, $R_{\star} = 10
  \rsun$, $b=0.995$, and $\log (\dot{M}/ \msunyr) = -4$. }
  \label{fig:Li}
\end{figure}

\citet{Li00} investigated the polarimetric variability of Wolf-Rayet
stars (WRs). Figure \ref{fig:Li} shows a comparison between their
results and the results of our model. Qualitatively the two studies
show a similar trend of steadily-falling \meanP\ with increasing
$\mathcal{N}$, behaviour also found by \citet{Richardson96}. However,
quantitatively the two differ by a factor of $\sim$5. The reason for
this is their different treatment of the clump density.

Li et al. fixed the clump angular size $\Delta\Omega$, clump thickness
$\Delta r$, and the electrons per clump $N_{\rm e}$ at constant
values. Therefore, the volume of the clump $V_{\rm cl}$ should
increase with the square of the distance, 

\begin{equation}
V_{\rm cl} \propto r^{2}
\label{equ:Li_V}
\end{equation}

They also stated that, following mass continuity, they scaled their
electron density as

\begin{equation}
n_{\rm e} \propto 1/(r^{2} v)
\label{equ:Li_mass}
\end{equation}

\noindent These statements are not consistent: if the number of
electrons per clump $N_{\rm e} = n_{\rm e} V_{\rm cl}$, then from
Eq.\,(\ref{equ:Li_mass}),

\begin{equation}
N_{\rm e} \propto 1/(n_{\rm e} v)
\label{equ:Li_Ne}
\end{equation}

\noindent Clearly, the number of electrons cannot be both constant
{\it and} proportional to 1/$v$. For the system of equations to be
self-consistent, the clumps must either,

\renewcommand{\labelenumi}{\roman{enumi})}
\begin{enumerate}
  \item be allowed to grow in thickness as they propagate through the
    wind, due to the differential acceleration between the front and
    back of the clumps; or
  \item have a fixed thickness and an {\it internal} clump density
    which scales as 1/$r^{2}$. The {\it space density} of the clumps
    still falls as 1/$r^{2} v$
\end{enumerate}
\renewcommand{\labelenumi}{\arabic{enumi}}

Li et al. integrated from $(x_{1} = x') \rightarrow (x_{2} = x'+\Delta
r$), where $\Delta r$ was constant, whilst allowing the electron
density to fall as 1/$r^{2} v$. They therefore only integrated over
the front $\Delta r$ of the clump, systematically underestimating the
polarization per clump. This explains their lower levels of \meanP\
compared to ours.

\begin{figure}[t]
\centering
\includegraphics[width=8.5cm]{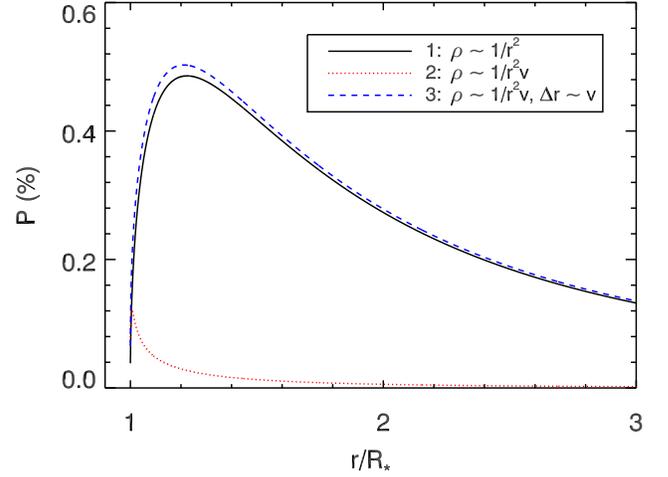}
  \caption[Comparison between 3 different scenarios for the
  calculation of polarization for one clump.]{Comparison between the
  three scenarios described in the text for calculating polarization
  as a function of distance for a single clump. Scenarios (1) and (3)
  are self-consistent and produce almost identical results, while (2)
  is different. Curve (2) peaks at $\sim$0.1\% extremely
  close to the star, while (1) and (3) peak at $\sim$0.5\% at a
  distance of 1.2$R_{\star}$. It is speculated that it is this factor
  of 5 difference in maximum polarization per clump which is
  responsible for the differing model results (see
  Fig.\,\ref{fig:Li}).  The curves are evaluated for $\tau = 1$,
  $\sin^{2}\chi =1$, $\Delta \Omega=0.04$, and $b=0.995$. }
  \label{fig:wrong}
\end{figure}


The differences in polarization per clump as a function of distance
from the star is illustrated in Fig.\,\ref{fig:wrong}. Three scenarios
are plotted:

\begin{enumerate}
\item Clump density $\propto 1/r^2$, with fixed $\Delta\Omega$,
  $\Delta r$.
\item Clump density $\propto 1/r^2 v$, with fixed $\Delta\Omega$,
  $\Delta r$.
\item Clump density $\propto 1/r^2 v$, with fixed $\Delta\Omega$, but
  with $\Delta r$ growing proportional to $v$.
\end{enumerate}

It can be seen that there is a significant difference between scenario
(2) and the self-consistent scenarios (1) and (3). Quantitatively, the
difference between the results of Li et al.~and this work can be
understood in terms of the maximum polarization per clump $P_{\rm
  max}$. In the limit where the ejection rate is much shorter than a
flow-time, i.e.~$\mathcal{N} \gg 1$, the time-averaged polarization
\meanP\ is proportional to $P_{\rm max}$. The maximum polarization per
clump in scenario (2) is $\sim$0.1\%, while for (1) and (3) $P_{\rm
  max} \sim 0.5\%$. This difference could explain the factor of
$\sim 5$ difference between this work and Li et al. for this set of
model parameters. 

It can also be seen from Fig.\,\ref{fig:wrong} that polarimetrically
there is little difference between scenarios (1) and (3). This is to
be expected, as both scenarios integrate over the same amount of
material and have the same optical depth. The minor difference when
$x$ is small is due to the finite star correction term.  The physical
differences between scenarios (1) and (3) are as follows: (1)
represents a thin shell of material being lifted off the surface of
the star; while (3) describes a localised {\it density
over-enhancement}, the material of which moves through the wind in the
same manner as the ambient material. We note that the two scenarios
have very different implications for the clump filling-factor $f_{\rm
cl}$: (3) has $f_{\rm cl}$ constant at all radii; while (1) has
$f_{\rm cl}$ which grows with distance from the star to some
asymptotic value. Figure \ref{fig:wrong} shows that while the two
scenarios differ in the physical situation they describe, the
resulting polarization is almost identical. As the computation times
for (1) are much shorter, and integrals easier, this scenario is
preferred throughout this study.

\subsection{Effect of various input parameters} \label{sec:params}
The free parameters in the model are the clump angular size $\Delta
\Omega$, ejection rate $\mathcal{N}$, and the stellar parameters
$\mu_{e}$, $R_{\star}$, $\dot{M}$, $v_{\infty}$, and $v_{0}$
(i.e. $b$). Of these, $\Delta \Omega$, $\mu_{e}$ and $\dot{M}$ will be fixed
throughout the parameter study, so the effect of these variables is
discussed below.

\begin{figure}[t]
  \centering
  \includegraphics[width=9cm,bb=45 60 708 510,clip]{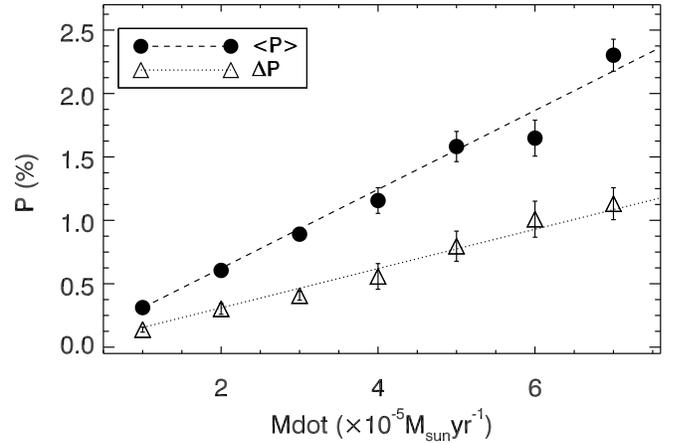}
  \caption{The effect of varying mass-loss rate on the time-averaged
  polarization and its variability in the optically-thin regime. The
  model parameters used are $v_{\infty}=180$km s$^{-1}$, $R_{\star} =
  100 \rsun$, $\Delta \Omega = 0.04$, $\Delta r = 0.01$. }
  \label{fig:mdot}
\end{figure}

\subsubsection{Mass-loss rate}
The effect of varying $\dot{M}$ for a fixed set of model parameters is
shown in Fig.~\ref{fig:mdot}. In the optically-thin limit, the
polarization per clump is directly proportional to $\dot{M}$
(Eqs.~\ref{equ:pshell2} and \ref{equ:tau0}). Figure \ref{fig:mdot}
shows, as expected, a linear relationship between $\dot{M}$ and
\meanP\ in the optically-thin regime, and a linear least-squares-fit
is overplotted to illustrate this. This linear relationship was also
found by \citet{Li00}. As the mass-loss rate increases the clumps
become optically-thick, therefore we would expect the trend to
flatten-off for extremely-high mass-loss rates.

Also shown in Fig.~\ref{fig:mdot} is the calculated polarimetric
variability, $\Delta P$. This is also linear with $\dot{M}$, with
$\Delta P /$\meanP\ consistently found to be $\sim 0.5 \rightarrow 0.6$,
not just as a function of mass-loss rate but for all simulations. This
can be understood as follows: if $P$ is truly random, one would expect
it to have the appearance of noise, with the amplitude $0 \rightarrow
P_{\rm max}$ to be roughly 3$\sigma$. The time-averaged value of $P$
will then be between these two at $\sim 1.5\sigma$. Therefore, the
quantity $\Delta P /$\meanP\ $\sim 1/1.5 \sim 0.67$.

\subsubsection{Clump size} \label{sec:results_size}
\begin{figure}[t]
  \centering
  \includegraphics[width=9cm,bb=45 60 708 510,clip]{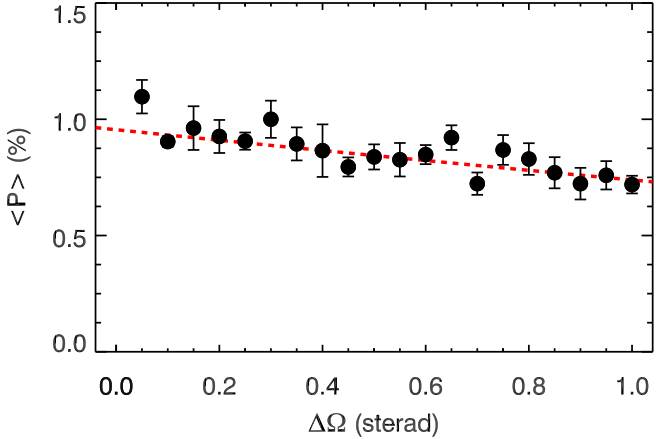}
  \caption{The effect of varying clump angular
  size on the time-averaged polarization. Overplotted is the relation
  in Eq.\,(\ref{equ:P_OM}), to which \meanP\ is expected to be
  proportional (dashed line). Errors on the data points are from the
  uncertainties in the means of seven independent simulations. Input
  parameters are the same as used in Figs.\,\ref{fig:Li} \&
  \ref{fig:mdot}.}
  \label{fig:OM}
\end{figure}

If the number of scatterers in a clump is conserved, then in theory
changing the clump's size should have little effect, as the same
amount of light will be scattered. However, there is a slight
trend of decreasing polarization for larger clumps: consider a `shell'
clump ejected perpendicular to the observer's line-of-sight. When
viewed on the plane of the sky, light scattering towards the observer
from the two opposing limbs of the clump will produce polarization
with slightly different position-angles. For a clump with a large
angular diameter, the position-angles of these two beams will tend
toward being perpendicular to each other, thus we expect less net
polarization from larger clumps. Quantitatively, for otherwise
constant input parameters, from inspection of Eqs. (\ref{equ:pshell2})
and (\ref{equ:tau0}) one would expect:

\begin{equation}
\langle P \rangle ~ \propto ~ (\mu_{2} - \mu_{2}^3)/\Delta \Omega
\label{equ:P_OM}
\end{equation}

\noindent where $\mu_{2} = 1 - \Delta \Omega / 2 \pi$. The effect of
varying $\Delta \Omega$ is shown in Fig.~\ref{fig:OM}, with the
relation in Eq.\,(\ref{equ:P_OM}) scaled and overplotted. The
behaviour of \meanP\ follows this relation well, with the scatter due to
the stochastic nature of the process. 

Large values of $\Delta \Omega$ ($\ga$0.5) are likely to be physically
unrealistic, and the fractional change in \meanP\ between $\Delta \Omega
= 0.01 \rightarrow 0.5$ is small (Fig.\,\ref{fig:OM}). Therefore, we
fix the angular size of the clumps to the value chosen by
\citet{Li00}, that of $\Delta \Omega = 0.04$.

%

\subsubsection{Mean mass-per electron}
The process considered in this study is electron-scattering, however
the electrons in each clump contribute a negligible amount to the
clump's mass. It is assumed that there are nuclei to account for each
of these free electrons, e.g. if the wind material was pure H then the
mass of a clump would be $N_{\rm e} m_{\rm H}$. However, the material
in the winds of massive evolved stars will have a non-negligible
content of heavier elements, so the mass per clump is determined by
the mean mass-per-electron, $\mu_{\rm e}$. From Eq. (\ref{equ:tau0})
it can be seen that the optical depth (and hence the polarization per
clump) is inversely proportional to $\mu_{\rm e}$. The reason for this
is that greater values of $\mu_{\rm e}$ mean fewer electrons in a
clump, reducing the electron-scattering optical depth.

The elemental abundances in a star's wind are determined to first
order by the chemistry of the natal ISM and the evolutionary phase of
the star. From recent abundance studies of hot supergiants it is
assumed that the elements heavier than helium contribute a negligible
amount to a clump's mass \citep[e.g.][]{Crowther06}. Determinations of
the helium number fraction $Y = n_{\rm He}/n_{\rm H}$ are not
straight-forward, due to subtle dependences of He line-strengths on
the details of radiative transfer calculations \citep[see
e.g. ][]{McErlean98}. Studies of LBVs have yielded values of $Y = 0.3
- 0.5$, \citep[][]{Crowther97,Najarro97}; while more recent studies of
OB supergiants with upgraded models suggest slightly lower values
\citep[$Y = 0.1 - 0.3$, ][]{Herrero02,Repolust05}. This implies a
value of $\mu_{\rm e} \sim 1.3 - 1.6$. As the uncertainty in the mean
mass-per-electron is small compared to that of other parameters, a
constant value of $\mu_{\rm e} = 1.5$ is adopted throughout the study.

\subsection{Polarization as a function of stellar radius}
\begin{figure}[t]
  \includegraphics[width=9cm,bb=40 0 550 720,clip]{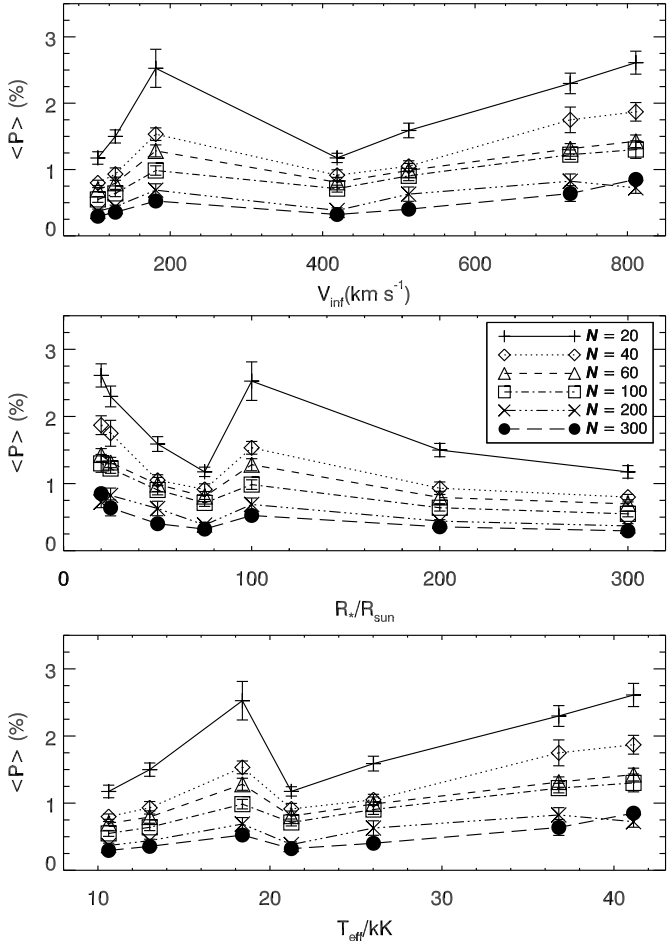}
  \caption{Time averaged polarization for
  constant values of $\mathcal{N}$, as a function of terminal wind
  speed ({\it top}), $R_{\star}$ ({\it middle}), and effective
  temperature ({\it bottom}). The stellar parameters are the same as
  those used throughout the parameter study, namely $L = 10^{6}
  \lsun$, $M = 30 \msun$, and $\log (\dot{M}/ \msunyr) = -4.5$. }
  \label{fig:N}
\end{figure}

Following the above arguments, this section investigates the
polarization produced by a clumpy wind across the range of stellar
radii inferred from LBV temperatures. For a given mass and constant
bolometric luminosity, the stellar radius and therefore effective
temperature and terminal velocity can be varied. Then, for a given
mass-loss rate, the clump ejection timescale and optical depth can be
varied, and the resulting polarization calculated. Values of $L =
10^{6} \lsun$, $M = 30 \msun$, and $\log (\dot{M}/ \msunyr) = -4.5$
are used, which are typical of the LBV \object{AG Car}
\citep{V-dK02}. The effect of varying $\dot{M}$ is significant, as it
is proportional to the time-averaged polarization (see previous
section). $L$ and $M$, in this parameterization, only affect the
terminal wind velocity, and so are redundant parameters if
$v_{\infty}$ is known. The polarization as a function of $v_{\infty}$,
$R_{\star}$ and $T_{\rm eff}$ for a range of ejection timescales
$\mathcal{N}$ is shown in Figure \ref{fig:N}.

\subsubsection{Varying ejection timescale}
If the mass-loss rate is conserved, increasing the number of clumps
ejected per wind flow-time $\mathcal{N}$ will decrease the optical
depth per clump. For low $\mathcal{N}$, the wind will consist of a
small number of dense clumps. For increasing $\mathcal{N}$, the wind
consists of a larger number of clumps with smaller optical depths,
until eventually the wind tends toward a smooth, spherically symmetric
outflow. Figure \ref{fig:N} illustrates this. It can be seen that
while the level of polarization decreases with increasing
$\mathcal{N}$, even at a high ejection rate there is still a
detectable level of polarization ($>$0.1\%) for the wind parameters used.

\subsubsection{Varying clump optical depth}
For constant clump size, decreasing the initial optical depth per
clump will also decrease the mass per clump. Therefore if the
mass-loss rate is to be conserved the ejection rate must {\it
increase}. The wind will be highly clumped at $\tau_{0} = 1$, and will
tend towards a smooth outflow at low values of $\tau$. For the model
parameters of Fig.\ \ref{fig:N}, $\mathcal{N}$ is linearly related to
$\tau$ via,

\begin{equation}
\tau = 3.6\times 10^{5} ~\mathcal{N}~ \displaystyle\left(
\frac{v_{\infty}}{\rm km\,s^{-1}} \right) ^{-1} \left(
\frac{R_{\star}} {R_{\odot}} \right) ^{-1}
\label{equ:tandn}
\end{equation}

\noindent Evaluating this expression for the lowest curves in
Fig.\,\ref{fig:N}, optical depths are in the region $\tau_{0} = 0.05
\rightarrow 0.1$. As $v_{\infty} \propto 1/\sqrt{R_{\star}}$
(Eq. \ref{equ:vinf_bsj}), optical depths are highest for small radius,
high terminal wind-speed stars.

\subsubsection{Overall behaviour of $P$ with $R_{\star}$}
It can be seen from Fig.\,\ref{fig:N} that, for a given ejection-rate
per wind flow-time there is a general trend of increasing polarization
with decreasing radius. This can be understood as follows: as the size
of the star is decreased, the clumps become smaller and denser,
producing more polarization per clump. Around the bi-stability zone,
the terminal velocity increases, and clumps spend less time close
to the star: the overall polarization drops again. 


The overall levels of polarization produced in the simulations are
high -- even at high $\mathcal{N} \sim 300$ where the wind is tending
toward a smooth outflow, observable residual levels of polarization
($\sim$0.3\%) are produced. This is discussed further in
Sect.\,\ref{sec:modeldisc}.




\subsection{Optically-thick clumps} \label{sec:resthick}
In this and the following section, the parameter-space is explored in
greater depth. To do this, the mass-loss rate, stellar radius and
terminal wind-speed are fixed at those of the LBV \object{P Cyg}, the
star for which most polarimetric observations exist. \citet{Hayes85},
\citet{Taylor91} and \citet{Nordsieck01} each present many individual
measurements, finding \meanP\ $\simeq 0.3 \pm 0.15 \%$. Indeed, it is
an extremely well studied object generally, with many independent
measurements of stellar / wind parameters in the literature. As the
star hasn't been significantly variable for $\ga$200 years, this means
that the input parameters of stellar radius, mass-loss rate and
terminal wind speed can be constrained. This then allows further
exploration of other details which may or may not be significant.

The parameters used are those derived by \citet{Najarro97}, namely
\begin{eqnarray}
R_{\star}/R_{\odot}  =  75, ~ \nonumber 
v_{\infty} = 185 {\rm km\,s^{-1}}, ~ \nonumber
\log ( \dot{M} / {\rm M_{\odot} yr^{-1}} ) = -4.52 \nonumber
\end{eqnarray}
\noindent These values were found through spectroscopic analysis with
non-LTE modelling. The model took no account of wind-clumping,
therefore the mass-loss rate may be an overestimate. Also, the authors
note that there is some degeneracy between $R_{\star}$, $v_{\infty}$
and $\dot{M}$ such that $\dot{M} / (v_{\infty} R_{\star}^{1.5}) \sim
{\it constant}$, but constrain $R_{\star}$ through other derived
values in the literature.

\begin{figure}[t]
\centering
  \includegraphics[width=9cm,bb=0 0 520 400,clip]{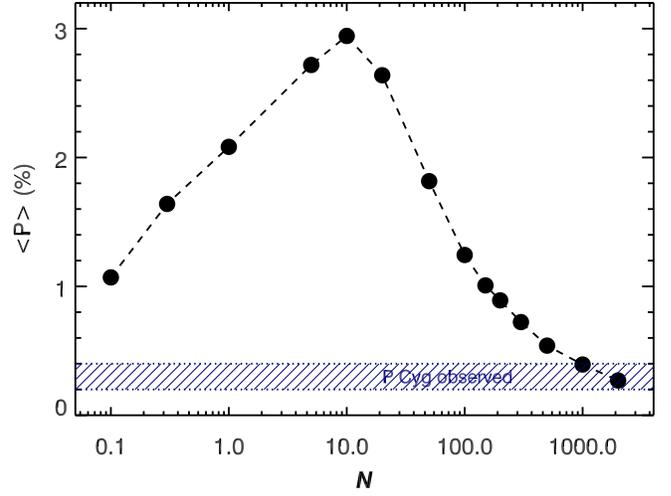}
  \caption[Polarization produced by optically thick
  clumps.]{Time-averaged polarization over a broad range of
  $\mathcal{N}$. At $\mathcal{N} \sim 20$, the optical depth per clump
  exceeds unity, and the overall polarization falls (see text for
  details). The P Cyg wind parameters derived by \citet{Najarro97} are
  used. The observed level of polarization for P Cyg is marked by the
  dash-dotted line.}
  \label{fig:pcyg_thick}
\end{figure}

Figure \ref{fig:pcyg_thick} shows the time-averaged polarization over
a broader range of $\mathcal{N}$ than in Fig.\ref{fig:N}. At high
$\mathcal{N}$, the static model was used to calculate $P$, and it can
be seen that the level of polarization observed for P Cyg is reached
at $\mathcal{N} \sim 2000$. 

At low $\mathcal{N}$, the density-per-clump becomes very high in order
to conserve the mass-loss rate, and exceeds unity below $\mathcal{N}
\sim 20$. At this point, the model switches to the optically-thick
approximation, that is the polarization per clump plateaus at $\tau >
1$. Therefore at $\mathcal{N} < 20$ the overall polarization begins to
fall again. We again point out that, in the optically-thick regime,
the polarization may become sensitive to clump morphology, and hence
the optically-thick part of the curve in Fig.\ \ref{fig:pcyg_thick}
{\it (left of $\mathcal{N} \simeq 10$)} may not be as quantitatively
robust as the thin-clump regime. However, the qualitative behaviour of
a turnover in \meanP\ is inevitable, as there is only a maximum amount
of polarization a single clump can produce. Therefore, we expect that
as clump density increases and the number of clumps becomes smaller,
there will be two solutions to match the observed level of
polarization -- the optically-thin, and optically-thick cases.

Extrapolating from Fig.\ \ref{fig:pcyg_thick}, the optically-thick
solution for P~Cyg is reached at at $\mathcal{N} \la 0.01$. As the
flow-time for P Cyg is about 3 days, this would imply ejection rates
of only a few per year even allowing for a factor of 10 difference in
the calculations.  This is deemed to be unlikely, for reasons which
will be discussed in greater detail in Sect.\,\ref{sec:modeldisc}, and
we therefore favour the optically-thin solution.

\begin{figure}[t]
\centering
  \includegraphics[width=9cm,bb=0 0 520 400,clip]{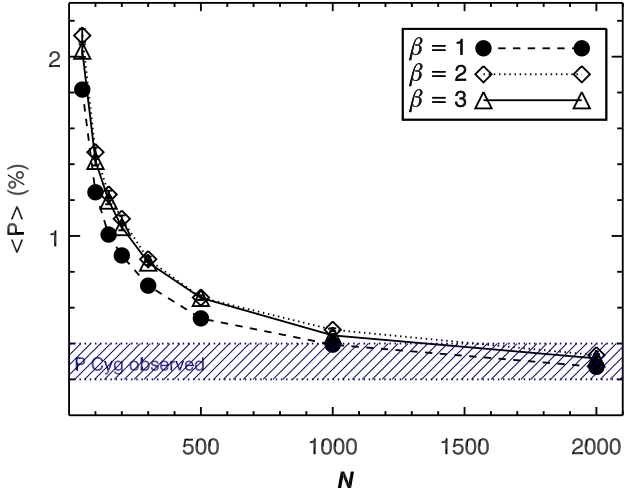}
  \caption[Polarization at higher wind acceleration]{The polarization
  produced when the wind acceleration parameter $\beta$ is increased,
  lowering the acceleration of the wind. Stellar parameters are the
  same as Fig.\,\ref{fig:pcyg_thick}. }
  \label{fig:pcyg_beta}
\end{figure}

\subsection{Lower wind acceleration}
In Fig.\,\ref{fig:pcyg_beta} the effect of increasing the wind
acceleration parameter $\beta$ is explored. It can be seen that, in
the case of P Cyg, increasing $\beta$ makes very little
difference. This is related to the ratio $v_{0}/v_{\infty}$: from
Fig.\,\ref{fig:oneclumpx}, it is clear that the greatest amount of
polarization is produced when the clump is within $\sim 6 R_{\star}$ of the
star. Figure \ref{fig:oneclumpt} shows that when $v_{0}/v_{\infty}$ is
relatively large, altering $\beta$ makes virtually no difference to
the dwell-time. Only when $v_{0}/v_{\infty}$ is small is the
dwell-time significantly affected by $\beta$. It is noted that at
larger $v_{\infty}$, increasing $\beta$ has a more pronounced effect,
as would be expect from the low $v_{0}/v_{\infty}$ case shown in
Fig.\,\ref{fig:oneclumpt}. At $v_{0}/v_{\infty} \sim 1/180$, the
polarization is increased by a factor of $\sim$1.5 when $\beta$ is
increased to 2.

\section{Discussion} \label{sec:modeldisc}

\subsection{A few massive clumps or many tiny clumps?}
In the approximation of a wind consisting of uniform clumps with no
inter-clump medium, the densities of the individual clumps need not be
large to produce the levels of polarization observed in massive
stars. Polarimetric variability on the scale observed in LBVs can in 
principle
be reproduced by two distinct regimes of $\mathcal{N}$: either
$\la$0.1 in the optically-thick case, or $\ga$1000 in the
optically-thin case (see Fig.\,\ref{fig:pcyg_thick}). 

In the former, the time between ejections would be of the order of
months (the flow-time of P Cyg is about 3 days), and one would expect
ejection events to be seen in the data of \citet{Hayes85} whose
temporal resolution was of order 1 day. Also, the density of the
clumps would be so large that the recombination emission would almost
certainly be conspicuous in time-resolved spectroscopy (quantitative
analysis of this is beyond the scope of the present work). For this
reason, the higher value of $\mathcal{N}$ is considered to be the most
likely value of the two.

At the inferred ejection rates of $\mathcal{N} \ga 10^3$, the inner
wind consists of thousands of low density clumps. In this regime, one
may expect the clumps to cancel each other out and leave a net
polarization of zero. However, only a slight imbalance is required to
produce residual polarization. For example, at $\mathcal{N}=2000$ and
the stellar parameters of P Cyg, the maximum polarization per clump is
0.006\%, and there are $\sim 7 \times 10^{3}$ clumps in the inner
2$R_{\star}$. A Poissonian 1$\sigma$ deviation from spherical symmetry
of $\sim$ 80 clumps is enough to produce a residual polarization of
 \meanP$ = 0.4\%$, which is the observed level of P~Cyg. Therefore, the
 polarization may simply result from the statistical deviations from
 spherical symmetry: {\it a slightly-fragmented wind} model.

 If observable levels of polarization are produced when a wind is so
 close to being homogeneous, it is reasonable to ask why intrinsic
 polarization is not present in all the stars observed by \citep{Davies05}. 
 The answer may be related to the star's mass-loss rate: as shown in
 Fig.\,\ref{fig:mdot}, the polarization produced by a given simulation
 is proportional to $\dot{M}$ if all other parameters are
 fixed. Therefore, if a star's ejection rate were fixed but its
 mass-loss rate were reduced from $10^{-4.5}$ to $10^{-6}\msunyr$, its
 average polarization could go from being as large as any observed to
 being completely undetectable. This suggests that intrinsic
 polarization may be ubiquitous among hot stars with mass-loss rates
 greater than $10^{-4.5}\msunyr$, as a measurable level of polarization
 can be produced from a statistical deviation from spherical symmetry.

 This conclusion is supported by the findings of \citet{Davies05}.
 Intrinsic polarization is more likely to be found in stars with
 strong H$\alpha$ emission; and for LBVs, whose terminal wind
 velocities are roughly the same, H$\alpha$ equivalent width may
 be considered to be a crude measure of mass-loss rate. It is also
 consistent with the lower incidence of polarization in O-supergiants
 (Osgs) and WRs, found in \citet{Harries98,Harries02}. Osgs have
 typical mass-loss rates of $\la 10^{-5}\msunyr$ \citep{Puls06}; and
 while WRs are considered to have $\dot{M} \sim 10^{-5} \rightarrow
 10^{-4}\,\msunyr$, ironically this may be reduced due to the effect
 of clumping \citep{H-K98}. The impact of clumping on LBV mass-loss
 rates is yet to be studied with the latest versions of model
 atmospheres.

 Of course, there is the possibility that the wind is a combination of
 the two scenarios, where a slightly fragmented wind produces
 low-level variability, combined with sporadic ejections of
 optically-thick clumps. Such a wind could be simulated by replacing
 the uniform clumps of the present model with clumps whose optical
 depths are determined by a power-law distribution. This would involve
 the introduction of an extra free parameter, the power-law index, and
 so it would be reasonable to look for observational evidence for such
 a scenario first before increasing the degenerate parameter
 space. The polarimetric signature of a dominant dense clump would be
 a jump in polarization and a PA which remained constant for several
 flow-times (few weeks for P Cyg), until its effect is `washed-out' by
 the smaller clumps close to the star. Also, it was shown by
 \citet{I-C04} that hierarchical clumping of ionized outflows can lead
 to atypical spectral indices in IR and radio free-free emission. Such
 spectra have slopes intermediate between the optically-thick/thin
 limits over several orders of magnitude in frequency. Hence, a
 combination of polarimetric monitoring and dynamic modelling, plus
 continuum radio observations, may be used to investigate hierarchical
 wind-clumping further.

\subsection{Wind `clumping factor'} \label{sec:fcl}
The clumping of hot star winds has been the subject of much recent
interest, as it has become apparent that a clumped wind can lead to
increased emission for the same amount of material. This means that
mass-loss rates derived from spectroscopic analysis under the
assumption of a smooth outflow are systematically over-estimated by
factors of 3 and higher \citep{Bouret05,Fullerton06,Puls06}. To
quantify this, the clumping of a star's wind is usually parameterized
by defining the {\it wind clumping factor} as a function of the
time-averaged wind density,

\begin{equation}
f_{\rm cl} = \frac{ \langle \rho \rangle ^{2} }{ \langle \rho^{2} \rangle }
\label{equ:f}
\end{equation}

\noindent in the limit where there is negligible inter-clump
medium. The clumping-factor itself is usually assumed to be a function
of wind-speed (and hence distance from star), 

\begin{equation}
f_{\rm cl} = f_{\infty} + ( 1 - f_{\infty} ) \exp(-v/v_{\rm cl})
\end{equation}

\noindent such that the wind starts off smooth, and clumping is
`switched on' at some scale-height above the photosphere when the velocity
reaches $v_{\rm cl}$. When incorporating clumping into ionization models,
some studies have chosen $v_{\rm cl}$ such that clumping becomes important
at large distances from the star \citep[e.g.][]{Dessart00}, where
hydrodynamical models have shown that clumping naturally arises from
instabilities in the line-driving mechanism \citep[][]{R-O02}. More
contemporary studies have chosen smaller values of $v_{\rm cl}$ to
initiate clumping just above the photosphere, based on the effect a
given clumping-law has on spectral line profiles
\citep{Hillier03}. More recently still, \citet{Puls06} presented an
investigation into the clumping factor of different regions of the
wind using simultaneous optical, IR and radio observations. They find
that, for stars with winds dense enough to produce line-emission, the
wind is strongly clumped in the inner 2$R_{\star}$, where the
line-emission is produced.

A direct determination of the clumping-law is not possible from
polarimetry alone. The polarization produced by a clump depends on the
amount of light the clump intercepts, and the thickness of the clump
is a redundant parameter provided the column-density remains
constant. Hence, polarimetry cannot distinguish between a dense,
flattened clump and a less-dense, extended clump, as both have the
same optical depth.

However, the aspects of wind-structure probed by polarimetry are
useful in conjuction with other observations in determining the
filling-factor of the wind. It has been shown here that polarimetry is
sensitive to the {\it number} of clumps that the wind is broken up
into. This may be relevant to other recent studies of wind-structure,
in particular the study of the X-ray line-profiles.

Luminous, O-type supergiants have been observed to emit X-ray
line-emission \citep[e.g. ][]{C-S83}, likely due to shocked gas
embedded in their outflowing winds \citep[][]{L-W80, OCR88}. In models
of such features the red-shifted emission from the back-hemisphere,
which has a larger path-length to travel to the observer, is predicted
to be attenuated by the intervening material, producing asymmetric
line profiles \citep{O-C01,I-G02}. However, this is not observed, with
the X-ray line profiles of Osgs remarkably symmetric in appearance
\citep[e.g.][]{Schultz00,W-C01}.

In order to explain this phenomenon, it has been suggested that either
the mass-loss rate must be reduced by a factor of $\ga5$
\citep{Cohen06}, or that the wind must be clumped in such a way as to
make the overall wind very {\it porous}, i.e. there must be large
separation between the clumps for the photons to leak through
\citep{O-C07}. The {\it porosity-length}, $h$, is defined as the ratio
of the clump size to the volume filling-factor, so having a large
value of $h$ sets the inter-clump distance to be very large compared
to the size of each clump. In this regime, the X-rays emitted from the
back hemisphere can travel through the wind relatively unattenuated,
and a symmetric line-profile is recovered.

Polarimetry can be used to constrain the available parameter-space of
these solutions. In the porous-wind model of \citet{O-C07}, there is
degeneracy between clump size and filling-factor -- that is, the wind
can consist of a few, large clumps with a large volume filling-factor;
or a large number of small, dense clumps and very small
filling-factor. These two regimes would produce very different
polarimetric signatures. In a wind containing very few clumps, the
polarimetric evolution of individual clumps could be traced, providing
there were only a handful of clumps in the inner $\sim 5
R_{\star}$. Resolving the dynamical evolution of individual clumps
would imply that the number of clumps was small, and we could then
place a lower-limit on the filling-factor for a given porosity-length.

If significant intrinsic polarization was detected, but no individual
clumps were resolved, this would allow us to place a lower-limit to
the number of clumps, and hence an upper-limit to $f$. Were no
intrinsic polarization to be observed at all, this would imply from
our models that the number of clumps so large that the wind was
tending towards a smooth outflow, which would not be consistent with
asymmetric X-ray line profiles. Given that \meanP\ scales with
$\dot{M}$ in our models, such a situation may instead be consistent
with a lower mass-loss rate for that object.

\section{Summary} \label{sec:modelconc}
We present an analytic investigation into the wind clumping of hot
stars, from the point-of-view of their polarimetric variability. In
particular, we study the effect of varying stellar parameters, with
the aim of reproducing the level and variability observed in LBVs.

The parameter space was defined such that the only key input parameter
for a given set of stellar/wind parameters was the ejection
timescale. It was found that, to quantitatively reproduce the
polarimetric variability of hot stars, the ejection timescale must
either be very short ($\ll$ 1 per hour) or very long ($\sim$ few per
year). In the latter case, the polarization arises from the sporadic
ejection of very dense clumps, and the variability timescale is very
long. In the former, the inner wind consists of thousands of
low-density clumps, with the observed polarization resulting from
random statistical deviations from spherical symmetry in a
slightly-fragmented wind. Here, significant changes are expected in
night-to-night monitoring. As the polarization of hot stars is
observed to be variable on very short timescales, and as they are not
associated with any significant spectroscopic variability that would
arise from dense clumps, the slightly-fragmented wind explanation is
favoured.

As polarization scales linearly with mass-loss rate, it is speculated
that all hot stars with high mass-loss rates are likely to display
polarimetric variations. This is consistent with the results of
\citep{Davies05} which reveal that intrinsic polarization is more
likely to be found in those LBVs with the strongest H$\alpha$
emission, which is a rough indicator of mass-loss rate. It is also
consistent with the lower incidence of polarization in O supergiants
and WRs, which likely have mass-loss rates factors of $\ga$5 lower.

\begin{acknowledgements}
We would like to thank the referee Rico Ignace and editor Steven Shore
for their inciteful suggestions and comments at the refereeing stage,
which improved the paper. We also thank John Brown, Qingkang Li, Stan
Owocki and Joachim Puls for fruitful discussions during the course of
this work. BD aknowledges funding by PPARC. JSV acknowledges RCUK for
his academic fellowship.
\end{acknowledgements}

 \bibliographystyle{aa}
 \bibliography{biblio}

\begin{thebibliography}{58}
\expandafter\ifx\csname natexlab\endcsname\relax\def\natexlab#1{#1}\fi

\bibitem[{{Bouret} {et~al.}(2005){Bouret}, {Lanz}, \& {Hillier}}]{Bouret05}
{Bouret}, J.-C., {Lanz}, T., \& {Hillier}, D.~J. 2005, \aap, 438, 301

\bibitem[{{Brown} {et~al.}(2000){Brown}, {Ignace}, \& {Cassinelli}}]{Brown00}
{Brown}, J.~C., {Ignace}, R., \& {Cassinelli}, J.~P. 2000, \aap, 356, 619

\bibitem[{{Brown} \& {McLean}(1977)}]{B-M77}
{Brown}, J.~C. \& {McLean}, I.~S. 1977, \aap, 57, 141

\bibitem[{Cassinelli {et~al.}(1987)Cassinelli, Nordsieck, \&
  Murison}]{Cassinelli87}
Cassinelli, J.~P., Nordsieck, K.~H., \& Murison, M.~A. 1987, ApJ, 317, 290

\bibitem[{{Cassinelli} \& {Swank}(1983)}]{C-S83}
{Cassinelli}, J.~P. \& {Swank}, J.~H. 1983, \apj, 271, 681

\bibitem[{{Code} \& {Whitney}(1995)}]{C-W95}
{Code}, A.~D. \& {Whitney}, B.~A. 1995, \apj, 441, 400

\bibitem[{{Cohen} {et~al.}(2006){Cohen}, {Leutenegger}, {Grizzard}, {Reed},
  {Kramer}, \& {Owocki}}]{Cohen06}
{Cohen}, D.~H., {Leutenegger}, M.~A., {Grizzard}, K.~T., {et~al.} 2006, \mnras,
  368, 1905

\bibitem[{{Crowther}(1997)}]{Crowther97}
{Crowther}, P.~A. 1997, in ASP Conf. Ser. 120: Luminous Blue Variables: Massive
  Stars in Transition, ed. A.~{Nota} \& H.~{Lamers}, 51--+

\bibitem[{{Crowther} {et~al.}(2006){Crowther}, {Lennon}, \&
  {Walborn}}]{Crowther06}
{Crowther}, P.~A., {Lennon}, D.~J., \& {Walborn}, N.~R. 2006, \aap, 446, 279

\bibitem[{{Davies} {et~al.}(2005){Davies}, {Oudmaijer}, \& {Vink}}]{Davies05}
{Davies}, B., {Oudmaijer}, R.~D., \& {Vink}, J.~S. 2005, A\&A, 439, 1107

\bibitem[{{Dessart} {et~al.}(2000){Dessart}, {Crowther}, {Hillier}, {Willis},
  {Morris}, \& {van der Hucht}}]{Dessart00}
{Dessart}, L., {Crowther}, P.~A., {Hillier}, D.~J., {et~al.} 2000, \mnras, 315,
  407

\bibitem[{{Fullerton} {et~al.}(2006){Fullerton}, {Massa}, \&
  {Prinja}}]{Fullerton06}
{Fullerton}, A.~W., {Massa}, D.~L., \& {Prinja}, R.~K. 2006, \apj, 637, 1025

\bibitem[{{Hamann} \& {Koesterke}(1998)}]{H-K98}
{Hamann}, W.-R. \& {Koesterke}, L. 1998, \aap, 335, 1003

\bibitem[{Harries(2000)}]{Harries00}
Harries, T.~J. 2000, MNRAS, 315, 722

\bibitem[{Harries {et~al.}(1998)Harries, Hillier, \& Howarth}]{Harries98}
Harries, T.~J., Hillier, D.~J., \& Howarth, I.~D. 1998, MNRAS, 296, 1072

\bibitem[{Harries {et~al.}(2002)Harries, Howarth, \& Evans}]{Harries02}
Harries, T.~J., Howarth, I.~D., \& Evans, C.~J. 2002, \mnras, 337, 341

\bibitem[{{Hayes}(1984{\natexlab{a}})}]{Hayes84}
{Hayes}, D.~P. 1984{\natexlab{a}}, \aj, 89, 1219

\bibitem[{{Hayes}(1984{\natexlab{b}})}]{Hayes84rsg}
{Hayes}, D.~P. 1984{\natexlab{b}}, \apjs, 55, 179

\bibitem[{{Hayes}(1985)}]{Hayes85}
{Hayes}, D.~P. 1985, \apj, 289, 726

\bibitem[{{Herrero} {et~al.}(2002){Herrero}, {Puls}, \& {Najarro}}]{Herrero02}
{Herrero}, A., {Puls}, J., \& {Najarro}, F. 2002, \aap, 396, 949

\bibitem[{{Hillier} {et~al.}(2003){Hillier}, {Lanz}, {Heap}, {Hubeny}, {Smith},
  {Evans}, {Lennon}, \& {Bouret}}]{Hillier03}
{Hillier}, D.~J., {Lanz}, T., {Heap}, S.~R., {et~al.} 2003, \apj, 588, 1039

\bibitem[{{Ignace} \& {Churchwell}(2004)}]{I-C04}
{Ignace}, R. \& {Churchwell}, E. 2004, \apj, 610, 351

\bibitem[{{Ignace} \& {Gayley}(2002)}]{I-G02}
{Ignace}, R. \& {Gayley}, K.~G. 2002, \apj, 568, 954

\bibitem[{{Kudritzki} \& {Puls}(2000)}]{K-P00}
{Kudritzki}, R.-P. \& {Puls}, J. 2000, \araa, 38, 613

\bibitem[{Lamers {et~al.}(1995)Lamers, Snow, \& Lindholm}]{Lamers95}
Lamers, H. J. G. L.~M., Snow, T.~P., \& Lindholm, D.~M. 1995, \apj, 455, 269

\bibitem[{{Li} {et~al.}(2000){Li}, {Brown}, {Ignace}, {Cassinelli}, \&
  {Oskinova}}]{Li00}
{Li}, Q., {Brown}, J.~C., {Ignace}, R., {Cassinelli}, J.~P., \& {Oskinova},
  L.~M. 2000, \aap, 357, 233

\bibitem[{{Lucy} \& {White}(1980)}]{L-W80}
{Lucy}, L.~B. \& {White}, R.~L. 1980, \apj, 241, 300

\bibitem[{{Lupie} \& {Nordsieck}(1987)}]{L-N87}
{Lupie}, O.~L. \& {Nordsieck}, K.~H. 1987, \aj, 93, 214

\bibitem[{{Machado} {et~al.}(2002){Machado}, {de Ara{\' u}jo}, {Pereira}, \&
  {Fernandes}}]{Machado02}
{Machado}, M.~A.~D., {de Ara{\' u}jo}, F.~X., {Pereira}, C.~B., \& {Fernandes},
  M.~B. 2002, \aap, 387, 151

\bibitem[{{McErlean} {et~al.}(1998){McErlean}, {Lennon}, \&
  {Dufton}}]{McErlean98}
{McErlean}, N.~D., {Lennon}, D.~J., \& {Dufton}, P.~L. 1998, \aap, 329, 613

\bibitem[{{Melgarejo} {et~al.}(2001){Melgarejo}, {Magalh{\~a}es}, {Carciofi},
  \& {Rodrigues}}]{Melgarejo01}
{Melgarejo}, R., {Magalh{\~a}es}, A.~M., {Carciofi}, A.~C., \& {Rodrigues},
  C.~V. 2001, \aap, 377, 581

\bibitem[{{Mokiem} {et~al.}(2007){Mokiem}, {de Koter}, {Vink}, \&
  {Puls}}]{mokiem07}
{Mokiem}, M., {de Koter}, A., {Vink}, J., \& {Puls}, J., e.~a. 2007, \aap

\bibitem[{{Najarro} {et~al.}(1997){Najarro}, {Hillier}, \& {Stahl}}]{Najarro97}
{Najarro}, F., {Hillier}, D.~J., \& {Stahl}, O. 1997, \aap, 326, 1117

\bibitem[{Nordsieck {et~al.}(2001)Nordsieck, Wisniewski, Babler, Meade,
  Anderson, Bjorkman, Code, Fox, Johnson, Weitenbeck, \& Zellner}]{Nordsieck01}
Nordsieck, K.~H., Wisniewski, J., Babler, B.~L., {et~al.} 2001, in P Cygni
  2000: 400 Years of Progress, ed. M.~de~Groot \& C.~Sterken (ASP Conference
  Series v.233)

\bibitem[{{Oskinova} {et~al.}(2006){Oskinova}, {Feldmeier}, \&
  {Hamann}}]{oskinova06}
{Oskinova}, L., {Feldmeier}, A., \& {Hamann}, W.-R. 2006, \aap, 372, 313

\bibitem[{{Owocki} \& {Cohen}(2001)}]{O-C01}
{Owocki}, S. \& {Cohen}, D. 2001, \apj, 559

\bibitem[{{Owocki} \& {Cohen}(2007)}]{O-C07}
{Owocki}, S. \& {Cohen}, D. 2007, \apj, in press

\bibitem[{{Owocki} {et~al.}(1988){Owocki}, {Castor}, \& {Rybicki}}]{OCR88}
{Owocki}, S.~P., {Castor}, J.~I., \& {Rybicki}, G.~B. 1988, \apj, 335, 914

\bibitem[{Pauldrach \& Puls(1990)}]{P-P90}
Pauldrach, A. W.~A. \& Puls, J. 1990, \aap, 237, 409

\bibitem[{{Puls} {et~al.}(2006){Puls}, {Markova}, {Scuderi}, {Stanghellini},
  {Taranova}, {Burnley}, \& {Howarth}}]{Puls06}
{Puls}, J., {Markova}, N., {Scuderi}, S., {et~al.} 2006, astro-ph/0604372

\bibitem[{{Repolust} {et~al.}(2005){Repolust}, {Puls}, {Hanson}, {Kudritzki},
  \& {Mokiem}}]{Repolust05}
{Repolust}, T., {Puls}, J., {Hanson}, M.~M., {Kudritzki}, R.-P., \& {Mokiem},
  M.~R. 2005, \aap, 440, 261

\bibitem[{{Repolust} {et~al.}(2004){Repolust}, {Puls}, \&
  {Herrero}}]{repolust04}
{Repolust}, T., {Puls}, J., \& {Herrero}, A. 2004, \aap, 415, 349

\bibitem[{{Richardson} {et~al.}(1996){Richardson}, {Brown}, \&
  {Simmons}}]{Richardson96}
{Richardson}, L.~L., {Brown}, J.~C., \& {Simmons}, J.~F.~L. 1996, \aap, 306,
  519

\bibitem[{{Robert} {et~al.}(1989){Robert}, {Moffat}, {Bastien}, {Drissen}, \&
  {St.-Louis}}]{Robert89}
{Robert}, C., {Moffat}, A.~F.~J., {Bastien}, P., {Drissen}, L., \& {St.-Louis},
  N. 1989, \apj, 347, 1034

\bibitem[{{Rodrigues} \& {Magalh{\~ a}es}(2000)}]{R-M00}
{Rodrigues}, C.~V. \& {Magalh{\~ a}es}, A.~M. 2000, \apj, 540, 412

\bibitem[{{Runacres} \& {Owocki}(2002)}]{R-O02}
{Runacres}, M.~C. \& {Owocki}, S.~P. 2002, \aap, 381, 1015

\bibitem[{Schulte-Ladbeck {et~al.}(1994)Schulte-Ladbeck, Clayton, Hillier,
  Harries, \& Howarth}]{S-L94}
Schulte-Ladbeck, R.~E., Clayton, G.~C., Hillier, D.~J., Harries, T.~J., \&
  Howarth, I.~D. 1994, ApJ, 429, 846

\bibitem[{Schulte-Ladbeck {et~al.}(1993)Schulte-Ladbeck, Leitherer, Clayton,
  Robert, Meade, Drissen, Nota, \& Schmutz}]{S-L93}
Schulte-Ladbeck, R.~E., Leitherer, C., Clayton, G.~C., {et~al.} 1993, \apj,
  407, 723

\bibitem[{{Schulz} {et~al.}(2000){Schulz}, {Canizares}, {Huenemoerder}, \&
  {Lee}}]{Schultz00}
{Schulz}, N.~S., {Canizares}, C.~R., {Huenemoerder}, D., \& {Lee}, J.~C. 2000,
  \apjl, 545, L135

\bibitem[{{St.-Louis} {et~al.}(1987){St.-Louis}, {Drissen}, {Moffat},
  {Bastien}, \& {Tapia}}]{StL87}
{St.-Louis}, N., {Drissen}, L., {Moffat}, A.~F.~J., {Bastien}, P., \& {Tapia},
  S. 1987, \apj, 322, 870

\bibitem[{Stahl {et~al.}(2001)Stahl, Jankovics, Kovács, Wolf, Schmutz, Kaufer,
  Rivinius, \& Szeifert}]{Stahl01}
Stahl, O., Jankovics, I., Kovács, J., {et~al.} 2001, \aap, 375, 54

\bibitem[{Taylor {et~al.}(1991)Taylor, Nordsieck, Schulte-Ladbeck, \&
  Bjorkman}]{Taylor91}
Taylor, M., Nordsieck, K.~H., Schulte-Ladbeck, R.~E., \& Bjorkman, K.~S. 1991,
  \aj, 102, 1197

\bibitem[{{van de Hulst}(1980)}]{vdHulst80}
{van de Hulst}, H.~C. 1980, {Multiple light scattering. Tables, formulas and
  applications} (New York: Academic Press, |c1980)

\bibitem[{{Vink} \& {de Koter}(2002)}]{V-dK02}
{Vink}, J.~S. \& {de Koter}, A. 2002, \aap, 393, 543

\bibitem[{Vink {et~al.}(1999)Vink, de~Koter, \& Lamers}]{Vink99}
Vink, J.~S., de~Koter, A., \& Lamers, H. J. G. L.~M. 1999, A\&A, 350, 181

\bibitem[{{Vink} {et~al.}(2000){Vink}, {de Koter}, \& {Lamers}}]{Vink00}
{Vink}, J.~S., {de Koter}, A., \& {Lamers}, H.~J.~G.~L.~M. 2000, \aap, 362, 295

\bibitem[{{Waldron} \& {Cassinelli}(2001)}]{W-C01}
{Waldron}, W.~L. \& {Cassinelli}, J.~P. 2001, \apjl, 548, L45

\bibitem[{{Zickgraf} \& {Schulte-Ladbeck}(1989)}]{Z-SL89}
{Zickgraf}, F.-J. \& {Schulte-Ladbeck}, R.~E. 1989, \aap, 214, 274

\end{thebibliography}

\end{document}